# BLOCH OSCILLATORS: MOTION OF WAVE-PACKETS

VINCENZO GRECCHI AND ANDREA SACCHETTI

ABSTRACT. In this paper we review some known results on the motion of Bloch Oscillators in the crystal momentum representation. We emphasize that the acceleration theorem, as usually stated by most of the authors, is incomplete, but in the case of pure Bloch states. In fact, the exact version of the acceleration theorem should contain a phase factor depending on both the time and crystal momentum variables. As we show this phase factor plays an essential role in order to understand the motion of Bloch Oscillators in the position variable.

## 1. INTRODUCTION

The motion of an electron in a periodic and regular crystal under the effect of an external electric field has been a subject of large interest since the early years of quantum mechanics [3, 24] (see also the recent paper [14] and the reference therein).

At present, the theoretical interest is mainly devoted to two different, even if related, problems.

One problem concerns the spectrum of the associated Hamiltonian

$$H_F = -\frac{\hbar^2}{2m}\frac{\partial^2}{\partial x^2} + V - Fx, \quad F = e\mathcal{E},$$

where, in the usual setting of the solid state physics, $m$ is the effective electron mass [1], $e$ is the electron charge, the electric field along the direction $x$ of the one dimensional crystal (or superlattice) has strength $\mathcal{E}$ and $V(x) = V(x+d)$ is the crystal potential with period $d$. Because of the periodicity of the crystal potential, the eigenvalues, if there, are displaced on regular ladders (usually called Wannier-Stark ladders). Wannier [22], in the sixties, predicted their existence in the framework of the perturbation theory. The very nature of these ladders was the object of a large debate, between Wannier himself, Zak and other authors (see [20] and the references therein), until the paper [2] which definitely proved that, at least for a regular periodic potential, the Hamiltonian $H_F$ has no discrete eigenvalues but, eventually, resonances. Just in these last years the existence of such resonances has been rigorously proved [9] and their imaginary part has been computed [11] (see also the review paper [8]).

The other problem concerns the time behavior of quantum observable, mainly the centroid of the wave-packet in both position and crystal momentum (called also quasi-momentum) representation. In fact, because of the periodicity of the

*Date*: October 7, 2005.
Notes of the talk given by Andrea Sacchetti at the workshop *Quantum dynamics and quantum transport*, University of Warwick, 6-10 September, 2004. This work is partially supported by the Italian MURST and INDAM-GNFM (project *Comportamenti Classici in Sistemi Quantistici*).

crystal potential, we expect to observe a periodic motion (Bloch Oscillator) of the wave-packet with period

$$t_B = 2\pi\hbar/Fd \tag{1}$$

and, surprisingly enough, an external homogeneous electric field produces, under some circumstances, a negative differential conductivity [7]. The naive theoretical justification of these facts is based on the acceleration theorem referred to pure Bloch states and on the analysis of the time behavior of a state initially prepared on a pure Bloch state of one band. In order to have a more realistic model for Bloch Oscillators and the negative differential conductivity, the recent theory of the miniband transport suggests to consider states that are initially distributed following a Boltzman-Maxwell type distribution (see [15] and the recent review paper [21]). Therefore, the version of the acceleration theorem referred to pure Bloch states appears to be inadequate in order to afford these new tasks. Not only, it also disagrees with some features of Bloch Oscillators recently observed [4, 19], namely the *breathing behaviour*, that is the wave-packet is spatially localized in a region which periodically enlarges and shrinks.

We emphasize also that, recently, the analysis of Bloch Oscillators, initially not fully localized in a region around the bottom of the energy, is of interest in the theory of ultracold atoms in periodical potentials [6].

The aim of this paper is twofold. At first, we review the acceleration theorem which is the corner-stone of the theory on the electronic transport in crystals; in particular, we show that it does not work for generic states because of a phase factor depending on the time and the crystal momentum variables. Then, collecting and improving our previous results [10, 11, 12, 13], we give a more powerful and complete version of the acceleration theorem and an expression for the centroid and the dispersion of the wave-packet in the position representation; as a result it follows the theoretical explanation of the *breathing motion* of Bloch Oscillators.

## 2. Centroid of the wave-packet in the Crystal Momentum Representation

The electron wave-packet $\psi(x,t)$ of an electron in a crystal superlattice under an homogenous electric field along the direction of the one dimesnional crystal satisfies to the time-dependent Schrödinger equation

$$i\hbar \frac{\partial}{\partial t}\psi = H_F \psi, \quad \psi = \psi(x,t) \tag{2}$$

where the wave-packet at the initial time $\psi^0(x) = \psi(x,0)$ satisfies to the normalization condition $\int_{-\infty}^{+\infty} |\psi^0(x)|^2 dx = 1$. Since $t_B$ is the actual unit of time it is more convenient to rescale the time as

$$\tau = \frac{Ft}{\hbar}$$

so that the Bloch period is given by

$$\tau_B = 2\pi$$

and the time-dependent Schrödinger equation takes the form

$$iF \frac{\partial}{\partial \tau}\psi = H_F \psi, \quad \psi = \psi(x,\tau) \tag{3}$$

The crystal momentum representation of the wave-packet is given by means of the periodic functions $\phi(k,\tau) = (\phi_n(k,\tau))_{n=1}^{\infty}$ defined as

$$\phi_n(k,\tau) = \int_{-\infty}^{+\infty} \psi(x,\tau)\bar{\varphi}_n(k,x)dx, \quad n=1,2,\ldots$$

and such that

(4) $$\psi(x,\tau) = \sum_{n=1}^{\infty} \int_{\mathcal{B}} \phi_n(k,\tau)\varphi_n(k,x)dk,$$

where $\varphi_n$ are the Bloch functions, $k$ is the crystal momentum variable belonging to the Brillouin zone $\mathcal{B} = [0, 2\pi/d)$. In such a representation equation (3) takes the form [5]

(5) $$\left[E_n(k) - iF\frac{\partial}{\partial k} - iF\frac{\partial}{\partial \tau}\right]\phi_n(k,\tau) - F\sum_{\ell=1}^{\infty} X_{n,\ell}(k)\phi_\ell(k,\tau) = 0$$

with initial condition

$$\phi_n^0(k) = \phi_n(k,0) = \int_{-\infty}^{+\infty} \psi^0(x)\bar{\varphi}_n(k,x)dx$$

where $E_n(k)$ are the band functions and where the coupling terms $X_{n,\ell}(k)$ are given by

$$X_{n,\ell}(k) = i\int_{-\infty}^{\infty} \bar{u}_n(k,x)\frac{\partial u_\ell(k,x)}{\partial k}dx$$

where

(6) $$\varphi_n(k,x) = e^{ikx}u_n(k,x), \quad u_n(k,x+d) = u_n(k,x).$$

The normalization of the wave-packet $\psi(x,\tau)$ implies that

$$\sum_{n=1}^{\infty} \int_{\mathcal{B}} |\phi_n(k,\tau)|^2 dk = \int_{-\infty}^{+\infty} |\psi(x,\tau)|^2 dx = 1$$

Following Callaway [5] we have that

**Theorem 1.** *Let*

$$\langle k \rangle^\tau = \langle \phi(k,\tau)|k|\phi(k,\tau)\rangle = \sum_{n=1}^{\infty} \int_{\mathcal{B}} k|\phi_n(k,\tau)|^2 dk.$$

*be the expectation value of the crystal momentum variable $k$ on the state $\phi$, then*

(7) $$\boxed{\langle k \rangle^\tau = \langle k \rangle^0 + \tau}$$

*where $\langle k \rangle^0$ is the expectation value of the crystal momentum variable on the initial state.*

The proof of this theorem is quite simple. Indeed let

$$(E\phi)_n = E_n\phi_n \quad \text{and} \quad (X\phi)_n = \sum_{\ell=1}^{\infty} X_{n,\ell}\phi_\ell,$$

where $X_{n,\ell} = \bar X_{\ell,n}$. From equation (5) it follows that the function

$$\Phi(k,\tau) = \sum_{n=1}^\infty \phi_n(k,\tau)$$

satisfies to the equation

$$\left[ F\frac{\partial}{\partial k} + F\frac{\partial}{\partial \tau} \right] \Phi(k,\tau) = 0$$

which has a solution of the type $\Phi(k,\tau) = u(k-\tau)$ where $u$ is an arbitrary periodic function of its argument such that

$$\int_\mathcal{B} |u(k)|^2 dk = \sum_{n=1}^\infty \int_\mathcal{B} |\phi_n(k,\tau)|^2 dk = 1.$$

Hence,

$$\begin{aligned}
\langle k \rangle^\tau &= \sum_{n=1}^\infty \int_\mathcal{B} k|\phi_n(k,\tau)|^2 dk \\
&= \int_\mathcal{B} k u(k-\tau) dk \\
&= \int_\mathcal{B} [k+\tau] u(k) dk \\
&= \langle k \rangle^0 + \tau
\end{aligned}$$

Equation (7) is often called *acceleration theorem*. We emphasize that we don't make any assumption on the initial wave-packet $\phi^0$, but the normalization condition.

## 3. Motion of the wave-packet in the crystal momentum representation

As a comment to the above theorem Callaway himself pointed out that: "One often find in the literature the alternative form $\frac{dk}{d\tau} = 1$ which must be understood as referring to the centroid of the packet" (see [5], pg. 468).

In fact, most of the authors of solid-state textbooks ascribe to the above acceleration theorem a different interpretation concerning the time behavior of the wave-packet. Thai is, they usually state that: "an electron which stays in a given state $k$ will *appear* to change its properties in terms of the states classified in $k$ at $\tau = 0$ as if $\frac{dk}{d\tau} = 1$. That is, an electron in a pure Bloch state at $\tau = 0$ will at a later time $\tau$ be in a state having the original $k$, but with all the other properties of the state originally at $k - \tau$." (see, e.g., [18], pg. 191).

We point out that the tunneling effect between bands is not considered in this interpretation; actually, this point is not really crucial since the tunneling time between bands is, typically, much larger than the period $\tau_B$ of Bloch Oscillators. Much more relevant is to remark that this interpretation only concerns pure Bloch states and not generic states prepared on one band; in fact, according with the criticism by Bouchard and Luban (see Appendix.3 in [4]), if the initial state is not a pure Bloch state then the above argument doesn't apply.

In order to overcome this flaw we state the following result (Theorem 2) which gives the behavior of Bloch Oscillators in the *decoupled band approximation* obtained by neglecting the interband interaction and where equation (5) takes the

form

(8) $$\left[E_n(k) - FX_{n,n}(k) - iF\frac{\partial}{\partial k} - iF\frac{\partial}{\partial \tau}\right]\phi_n(k,\tau) = 0, \ n = 1, 2, \ldots$$

**Theorem 2.** *The solution $\phi_n(k,\tau)$ of the decoupled band approximation (8) satisfying the initial condition $\phi_n(k,0) = \phi_n^0(k)$ at $\tau = 0$ is given by*

(9) $$\boxed{\phi_n(k,\tau) = e^{i\theta_n(k,\tau)}\phi_n^0(k-\tau)}$$

*where*

(10) $$\boxed{\theta_n(k,\tau) = -\frac{1}{F}\int_{k-\tau}^{k}[E_n(q) - FX_{n,n}(q)]\,dq}$$

*is a phase factor.*

In order to prove (9) and (10) we set $\xi = k - \tau$, then equation (8) takes the form

$$\left[E_n(k) - FX_{n,n}(k) - iF\frac{\partial}{\partial k}\right]\phi_n(k,\xi) = 0, \ n = 1, 2, \ldots$$

which has solution

$$\phi_n(k,\xi) = h_n(\xi)\exp\left[-\frac{i}{F}\int_0^k[E_n(q) - FX_{n,n}(q)]\,dq\right]$$

where $h_n(\xi)$ is an arbitrary function of its argument. In order to have $\phi_n^0$ at the time $\tau = 0$, that is at $\xi = k$, then

$$h_n(k) = \phi_n^0(k)\exp\left[\frac{i}{F}\int_0^k[E_n(q) - FX_{n,n}(q)]\,dq\right]$$

obtaining (9) and (10).

*Remark 3.* From (9) it follows that the time behavior of Bloch Oscillators in the crystal momentum representation is given by means of an uniform translation $k \to k - \tau$ (as usually expected) together with a change of the phase.

*Remark 4.* We underline that our result agrees with the previous one (7). In particular, from (9) it follows that

$$\begin{aligned}\langle k \rangle^\tau &= \sum_{n=1}^\infty \int_\mathcal{B} k|\phi_n(k,\tau)|^2 dk \\ &= \sum_{n=1}^\infty \int_\mathcal{B} k|\phi_n^0(k-\tau)|^2 dk \\ &= \sum_{n=1}^\infty \int_\mathcal{B} [k+\tau]|\phi_n^0(k)|^2 dk \\ &= \langle k \rangle^0 + \tau\end{aligned}$$

obtaining again (7).

If the initial state $\phi^0$ is a *pure Bloch state* then, as also proved by Houston [16] and Wannier (see equation (45) in [23]), $\phi(k,\tau) = \phi^0(k-\tau)e^{i\theta_n(k_0,\tau)}$ coincides, up to a phase factor that does not depend on $k$, with the Bloch state having quasimomentum satisfying the equation $\frac{dk}{d\tau} = 1$, in agreement with the previous interpretation. We emphasize that, in such a case, the phase factor $\theta_n(k_0,\tau)$ does not play any particular role since it does not depend on $k$. In contrast, if the initial state $\phi^0$ *does not coincide with a pure Bloch state*, the phase factor appearing in (9), which actually depends on $k$, has no previously considered by other authors for what know by us; as we'll show later it plays a crucial role in order to obtain the dynamics of the wave-packet $\psi(x,\tau)$ in the position representation.

It is clear that the crucial point is the validity of the decoupled approximation (8) for times of the order of the Bloch period $t_B$. In the limit of small electric field the tunneling time is much larger than the Bloch period; then we expect that the tunneling effect between different bands should be negligible and that the dominant term of the wave-packet in the crystal momentum representation is given by equation (9). However, the rigorous proof of this fact is still lacking, but in the case of a crystal potential with a finite number of open gaps [12].

4. EXPECTATION VALUE OF THE POSITION OPERATOR

Going back to the position representation, from Theorem 2, Remark 6 and equation (4) it follows that the solution of the time-dependent Schrödinger equation (3) has dominant term given by

$$(11) \qquad \psi(x,\tau) = \sum_{n=1}^{\infty} \int_{\mathcal{B}} \phi_n^0(k-\tau) e^{i\theta_n(k,\tau)} \varphi_n(k,x) dk$$

for times of the order of the Bloch period $\tau_B$ and in the limit of small electric field.

Here, from (11), we give the expectation value of the position operator and the variance.

**Theorem 5.** *Let*

$$\langle x \rangle^\tau = \langle \psi(x,\tau) | x | \psi(x,\tau) \rangle = \int_{-\infty}^{+\infty} x |\psi(x,\tau)|^2 dx$$

*be the expectation value of the position observable and let $S^\tau = \left\langle [x - \langle x \rangle^\tau]^2 \right\rangle^\tau$ its variance. Then, in the limit of small electric field and for times of the order of the Bloch period, the dominant terms are given by*

$$(12) \qquad \langle x \rangle^\tau - \langle x \rangle^0 = \sum_{n=1}^{+\infty} \frac{1}{F} \int_{\mathcal{B}} \left| \phi_n^0(k) \right|^2 \left[ E_n(k+\tau) - E_n(k) \right] dk$$

*and, if $\phi^0$ is a real-valued function, then*

$$S^\tau - S^0 = -\left( [\langle x \rangle^\tau]^2 - [\langle x \rangle^0]^2 \right) +$$

$$(13) \qquad\qquad + \frac{1}{F^2} \sum_{n=1}^{\infty} \int_{\mathcal{B}} [E_n(k+\tau) - E_n(k)]^2 |\phi_n^0(k)|^2 dk.$$

Here, following [13], we prove equations (12) and (13) making use of the acceleration theorem for the motion of wave-packet in the crystal momentum representation. In fact, a different proof, under some technical assumptions on the external field, has been recently given by [17].

From equations (3) and (4), from the acceleration theorem in the form (9) and from Remark 6 then formally follows that

$$\begin{aligned} Fx\psi &= -iF\frac{\partial \psi}{\partial \tau} + \left[-\frac{\hbar^2}{2m}\frac{\partial^2}{\partial x^2} + V\right]\psi \\ &= \sum_{n=1}^{\infty} \int_{\mathcal{B}} d_n(k,\tau)\varphi_n(k,x)dk \end{aligned}$$

where

$$\begin{aligned} d_n(k,\tau) &= -iF\frac{\partial \phi_n(k,\tau)}{\partial \tau} + E_n(k)\phi_n(k,\tau) \\ &= \left[iF\frac{\partial \phi_n^0(k-\tau)}{\partial k} + (E_n(k) - E_n(k-\tau))\phi_n^0(k-\tau)\right] e^{i\theta_n(k,\tau)}. \end{aligned}$$

From this fact and since $\int_{-\infty}^{+\infty} \bar{\varphi}_m(x,k')\varphi_n(x,k)dx = \delta(k-k')\delta_n^m$ then it follows that

$$\begin{aligned} F\langle x\rangle^\tau &= \int_{-\infty}^{+\infty} \bar{\psi}(x,\tau)Fx\psi(x,\tau)dx \\ &= \sum_{n,m=1}^{\infty} \int_{-\infty}^{+\infty} \int_{\mathcal{B}}\int_{\mathcal{B}} \bar{\phi}_m(k',\tau)d_n(k,\tau)\bar{\varphi}_m(k',x)\varphi_n(k,x)dkdk'dx \\ &= \sum_{n=1}^{\infty} \int_{\mathcal{B}} \bar{\phi}_n(k,\tau)d_n(k,\tau)dk \\ &= g(\tau) + \sum_{n=1}^{\infty} \int_{\mathcal{B}} [E_n(k) - E_n(k-\tau)]\,|\phi_n^0(k-\tau)|^2 dk \\ &= g(0) + \sum_{n=1}^{\infty} \int_{\mathcal{B}} [E_n(k+\tau) - E_n(k)]\,|\phi_n^0(k)|^2 dk \end{aligned}$$

where the term

$$\begin{aligned} g(\tau) &= \sum_{n=1}^{\infty} iF \int_{\mathcal{B}} \bar{\phi}_n^0(k-\tau)\frac{\partial \phi_n^0(k-\tau)}{\partial k}dk \\ &= \sum_{n=1}^{\infty} iF \int_{\mathcal{B}} \bar{\phi}_n^0(k)\frac{\partial \phi_n^0(k)}{\partial k}dk = g(0) \end{aligned}$$

is independent of time and, from the previous equation, it coincides with $F\langle x\rangle^0$. Hence (12) follows.

In order to compute the variance $S^\tau = \left\langle [x - \langle x\rangle^\tau]^2\right\rangle^\tau = \left\langle x^2\right\rangle^\tau - [\langle x\rangle^\tau]^2$ we make use of the above equation where, assuming that the function $\phi^0$ is real-valued, we

obtain that

$$
\begin{aligned}
F^2 \langle x^2 \rangle^\tau &= \langle Fx\psi, Fx\psi \rangle \\
&= \left\langle \sum_{m=1}^\infty \int_{\mathcal{B}} d_m(k',\tau)\varphi_m(k',x)dk', \sum_{n=1}^\infty \int_{\mathcal{B}} d_n(k,\tau)\varphi_n(k,x)dk \right\rangle \\
&= \sum_{n=1}^\infty \int_{\mathcal{B}} |d_n(k,\tau)|^2 dk \\
&= \sum_{n=1}^\infty \int_{\mathcal{B}} |d_n(k+\tau,\tau)|^2 dk \\
&= \sum_{n=1}^\infty \int_{\mathcal{B}} \left| iF \frac{\partial \phi_n^0(k)}{\partial k} + [E_n(k+\tau) - E_n(k)] \phi_n^0(k) \right|^2 dk \\
&= F^2 \langle x^2 \rangle^0 + \sum_{n=1}^\infty \int_{\mathcal{B}} [E_n(k+\tau) - E_n(k)]^2 |\phi_n^0(k)|^2 dk
\end{aligned}
$$

where

$$
\langle x^2 \rangle^0 = \sum_{n=1}^\infty \int_{\mathcal{B}} \left| \frac{\partial \phi_n^0(k)}{\partial k} \right|^2 dk
$$

proving so the theorem.

*Remark 6.* Our result extend the one obtained by Bouchard and Louban (see eq. (16) in [4]) for the special case of a state initially prepared on the first band, where they assumed that the initial state $\phi_1^0$ is independent of the quasi-momentum $k$ and where the first band is given by a cosine function.

Equation (12) can be also written as

(14) $$\boxed{F\left[\langle x \rangle^\tau - \langle x \rangle^0\right] = \langle E(k+\tau) \rangle^0 - \langle E(k) \rangle^0}$$

where in the left hand side the mean value has to be intended in the position variable while in the right hand side the mean value has to be intended in the crystal momentum variable. For what concerns the variance we remark that, if $\phi^0$ is a real-valued function then the function $\psi^0$ has the property of symmetry $|\psi^0(-x)| = |\psi^0(x)|$; thus $\langle x \rangle^0 = 0$ and equation (13) can be written as

$$\boxed{F^2(S^\tau - S^0) = \left\langle [E(k+\tau) - E(k)]^2 \right\rangle^0 - \left[\langle E(k+\tau) - E(k) \rangle^0\right]^2}$$

where

$$
S^0 = \sum_{n=1}^\infty \int_{\mathcal{B}} \left| \frac{\partial \phi_n^0(k)}{\partial k} \right|^2 dk.
$$

Furthermore, from (14), it is easy to show that the classical results on the mean velocity and the effective mass [5] can be recovered:

$$
\frac{d\langle x \rangle^t}{dt} = \frac{1}{\hbar} \left\langle \frac{\partial E(k+\tau)}{\partial k} \right\rangle^0
$$

and

$$\frac{d^2 \langle x \rangle^t}{dt^2} = \frac{F}{\hbar^2} \left\langle \frac{\partial^2 E(k+\tau)}{\partial k^2} \right\rangle^0.$$

## 5. Motion of the wave-packet in the position representation: breathing behavior versus soliton like shape

Now, making use of the previous results we explain the motion of Bloch Oscillators, in full agreement with numerical and experimental results (see, e.g., [4, 19]). To this end we consider the motion of Bloch Oscillators initially prepared on the first band; in such a case, the dominant term (that is up to terms of higher orders in $F$) of the wavefunction is given by means of formulas (6), (9) and (11)

$$(15) \quad \psi(\chi,\tau) = \int_{\mathcal{B}} \phi_1^0(k-\tau) e^{\frac{i}{F}\left[k\chi - \int_{k-\tau}^{k}[E_1(q) - FX_{1,1}(q)]dq\right]} u_1(k,\chi/F) dk$$

where we rescale the position variable as $\chi = Fx$.

In the limit of small $F$ the stationary phase theorem gives that the wavefunction $\psi(\chi,\tau)$ is localized on the interval $[-\chi(\tau), \chi(\tau)]$ depending on time, where

$$\chi(\tau) = \max_{k \in \mathcal{B}} [E_1(k) - E_1(k-\tau)],$$

and it is exponentially vanishing outside. Indeed, stationary points are the real solutions $k$ of the equation $E_1(k) - E_1(k-\tau) + \chi = 0$. The function $\chi(\tau)$ is a periodic function, with period $\tau_B$, has minimum value 0 at $\tau = 0$ and has maximum value $\Delta$ at $\tau = \frac{1}{2}\tau_B$, where $\Delta$ is the width of the first band. Hence, Bloch Oscillators, represented by wave-functions $\psi(x,\tau)$, perform a *breathing motion* since the interval of localization $[-\chi(\tau)/F, \chi(\tau)/F]$ periodically enlarges and shrinks; in particular, the maximum interval of localization has width $2\Delta/F$, in agreement with the tilted band Zener picture (see Figure 1).

In order to better understand the motion of Bloch Oscillators inside the interval of localization we compute the centroid $\langle x \rangle^\tau$ and the standard deviation $\sigma^\tau = \sqrt{S^\tau}$ of the wave packet for different initial states. Hereafter, for the sake of simplicity, we assume that the first band function is simply given by $E_1(k) = \frac{1}{2}\Delta[1 - \cos(kd)]$, where $\Delta$ is the width of the first band, and that the period of the crystal is $d = 1$.

In the first case we consider a state that initially coincides with an exact Wannier states where $\phi_1^0(k) = \sqrt{d/2\pi}$ (Figure 2, dots line), that is the electron wave-packet is initially localized on one site of the superlattice. In such a case the wave-packet exhibits a symmetrical motion and the centroid remains fixed (Figure 3, dots line). As appears from Figure 4 (dots line) the Bloch Oscillator performs an actual breathing motion, that is the wave-packet, initially localized on a single site of the superlattice, periodically enlarges (for increasing variance) and shrinks (for decreasing variance) without moving its center. As a result, we expect to have no electronic current, according to the fact that there is no conductance in full bands [21].

In the second case we consider the opposite situation, where the state is initially localized in the crystal momentum around $k_0 = 0$ according with the following distribution

$$(16) \qquad \phi_1^0(k) = ce^{-k^2/2\rho^2}, \quad c = \left[\sqrt{\pi}\rho \operatorname{erf}\left(\frac{\pi}{\rho d}\right)\right]^{-1/2}$$

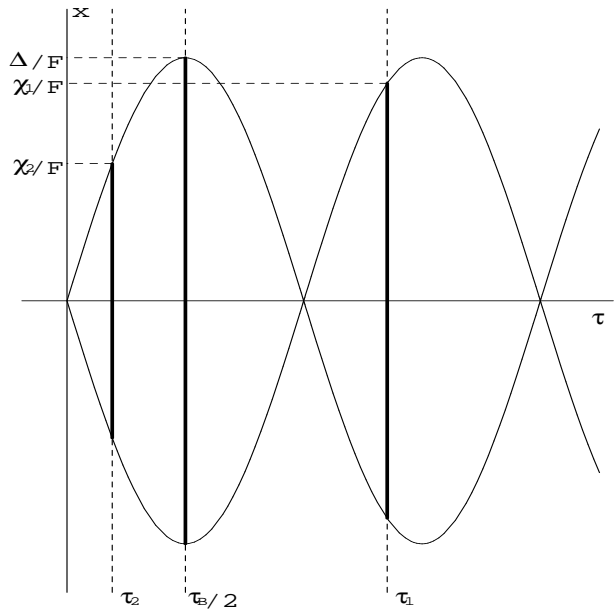

FIGURE 1. The Bloch Oscillator, represented by the wave-function $\psi(x,\tau)$, is localized inside the interval $[-\chi(\tau)/F, \chi(\tau)/F]$ (bold vertical lines) and it is exponentially vanishing outside this interval. Since $\chi(\tau)$ is a positive periodic function, with period $\tau_B$, then the *breathing behaviour* of Bloch Oscillators clearly appears. Only at $\tau = \frac{1}{2}\tau_B$ the interval of localization coincides with the interval $[-\Delta/F, \Delta/F]$, where $\Delta$ is the width of the first band, as predicted by means of the Zener tilted band picture.

for $\rho > 0$ (Figure 2, broken and full lines for, respectively, $\rho = 1$ and $\rho = 0.1$), and periodically arranged on the Brillouin zone. In such a case the motion of the wave-packet is similar to the one of a soliton (a soliton is, roughly speaking, a wave-packet that translates in space maintaining its initial shape); indeed, if $\rho$ is small enough, then the centroid of the wave-packet moves forward for a length $\Delta/F$ maintaining a well localized shape and then it goes back to the initial position (see full lines in Figures 3, 4). In particular, in the limit case of $\rho \to 0$ then we obtain a pure Bloch state at $k_0$ and the centroid of the wave-packet moves according with equation $F\left[\langle x \rangle^\tau - \langle x \rangle^0\right] = [E_1(k_0 + \tau) - E_1(k_0)]$.

## 6. Conclusions

In conclusion, here we have shown that the time behavior of a Bloch Oscillator cannot be fully explained by means of the acceleration theorem (7), but in the case of a pure Bloch state. In the improved and exact version of the acceleration theorem (9) a phase factor should be taken into account, the relevance of this term for the motion of the wave-packet in the position representation is explicitely shown.

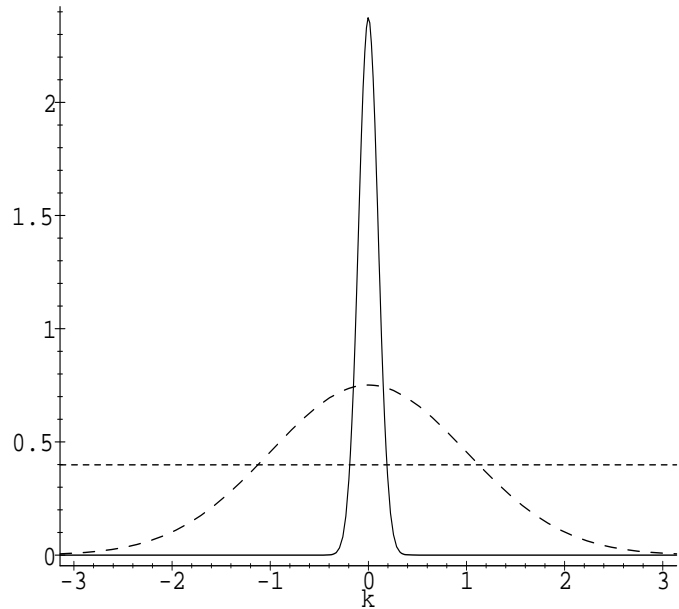

FIGURE 2. We plot the amplitude $|\phi_1^0(k)|^2$ for different initial wave-packets. Dots line represent the Wannier state initially localized on one single cell of the superlattice; broken and full lines, respectively, represent Gaussian like wave-packets with $\rho = 1$ and $\rho = 0.1$.

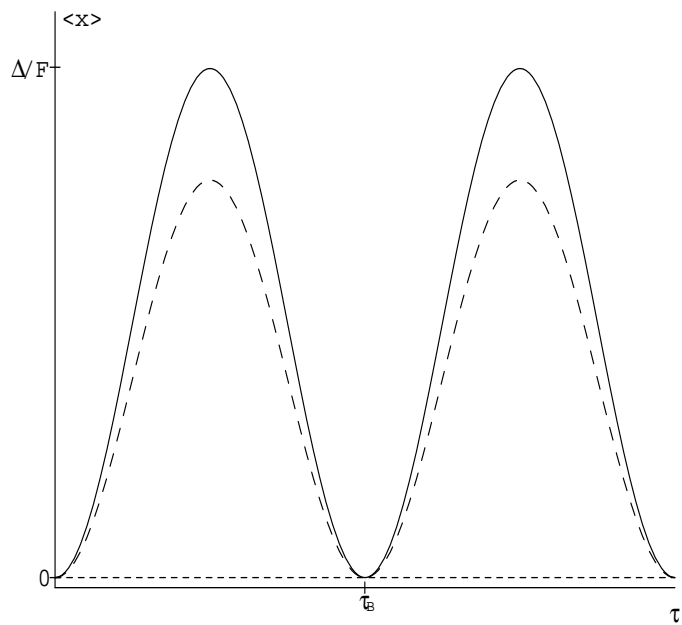

FIGURE 3. We plot the function $\langle x \rangle^\tau$, which represents the motion of the centroid of the wave-packet, $\Delta$ is the width of the first band. For a state initially prepared on a Wannier state then we don't have motion of the centroid (dot line). In contrast, as much as the state is initially well localized in the crystal momentum variable then the centroid of the wave-packet perform a full motion inside the band (broken and full lines).

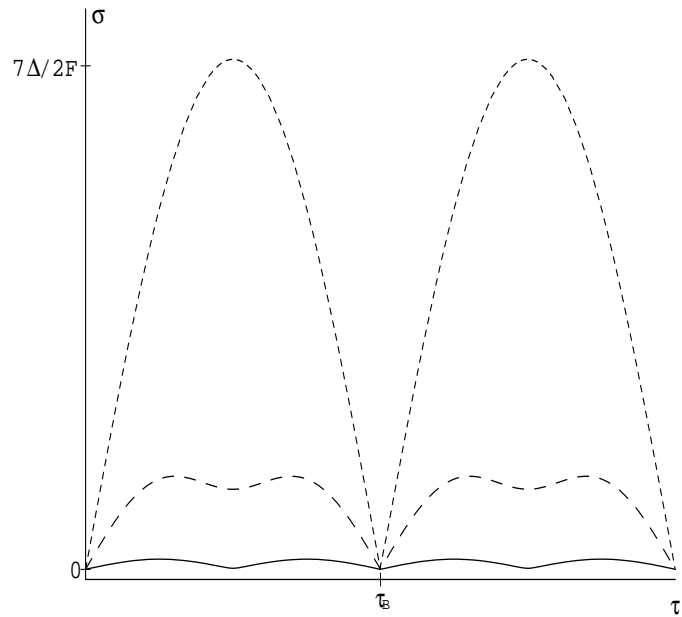

FIGURE 4. We plot the standard deviation $\sigma^\tau$ of the wave-packet for Wannier initial state (dot line) and Gaussian type states with (broken and full lines). In the first case we have a breathing behavior of the motion where the wave-packet periodically enlarges and shrinks. In the latter cases the wave-packet remains well localized during the motion of its centroid as for soliton-like wave-packet

Dipartimento di Matematica, Università di Bologna, Piazza di Porta S. Donato 5, I–40127 Bologna, Italy, email: Grecchi@dm.unibo.it.

Dipartimento di Matematica, Università di Modena e Reggio Emilia, Via Campi 213/B, I–41100 Modena, Italy, email: Sacchetti@unimo.it